\documentclass[letterpaper, 10 pt, conference]{ieeeconf}

\IEEEoverridecommandlockouts

\usepackage{cite}
\usepackage{amsmath,amssymb,amsfonts}
\usepackage{algorithmic}
\usepackage{graphicx}
\usepackage{textcomp}
\usepackage{xcolor}
\usepackage{siunitx}
\usepackage{float}
\usepackage[symbol]{footmisc}
\usepackage{ar}
\usepackage{balance}

\newcommand{\bs}[1]{\boldsymbol{#1}}  
\newcommand{\ts}[1]{\text{#1}} 

\renewcommand{\baselinestretch}{0.92}

\def\BibTeX{{\rm B\kern-.05em{\sc i\kern-.025em b}\kern-.08em

    T\kern-.1667em\lower.7ex\hbox{E}\kern-.125emX}}

\begin{document}

\title{\LARGE \bf VLEIBot: A New $\boldsymbol{45}$-mg Swimming Microrobot Driven by a Bioinspired Anguilliform Propulsor \\

\thanks{This work was partially funded by the Washington State University (WSU) Foundation and the Palouse Club through a Cougar Cage Award to \mbox{N.\,O.\,P\'erez-Arancibia}. Additional funding was provided by the WSU Voiland College of Engineering and Architecture through a start-up fund to \mbox{N.\,O.\,P\'erez-Arancibia}.} %
\thanks{E.~K.~Blankenship and C.~K.~Trygstad contributed equally to this work.}
\thanks{The authors are with the School of Mechanical and Materials Engineering, Washington State University (WSU), Pullman,\,WA\,99164,\,USA. Corresponding authors' \mbox{e-mail:} 
{\tt conor.trygstad@wsu.edu}~(C.\,K.\,T.);
{\tt n.perezarancibia@wsu.edu} (N.\,O.\,P.-A.).}%
}
\author{Elijah K. Blankenship, Conor K. Trygstad, Francisco M. F. R. Gonçalves, and N\'estor O. P\'erez-Arancibia}

\maketitle
\thispagestyle{empty}
\pagestyle{empty}

\begin{abstract}
This paper presents the VLEIBot\footnote[1]{We pronounce this word as \textit{\textbf{vl-ahy}-bot}, thus producing the same sound in the pronunciations of \textit{eigenvalue} and \textit{Einstein}.} (\textit{Very Little Eel-Inspired roBot}), a \mbox{$\bs{45}$-mg/$\bs{23}$-$\ts{mm}^{\bs{3}}$} microrobotic swimmer that is propelled by a bioinspired anguilliform propulsor. The propulsor is excited by a single \mbox{$\bs{6}$-mg} \mbox{\textit{high-work-density}} (HWD) microactuator and undulates periodically due to wave propagation phenomena generated by \textit{\mbox{fluid-structure interaction}} (FSI) during swimming. The microactuator is composed of a \mbox{carbon-fiber} beam, which functions as a leaf spring, and \textit{shape-memory alloy} (SMA) wires, which deform cyclically when excited periodically using Joule heating. The \mbox{VLEIBot} can swim at speeds as high as \mbox{$\bs{15.1}\,\ts{mm} \cdot \ts{s}^{\bs{-1}}$ ($\bs{0.33}\,\ts{Bl} \cdot \ts{s}^{\bs{-1}}$)} when driven with a \mbox{heuristically-optimized} propulsor. To improve maneuverability, we evolved the \mbox{VLEIBot} design into the \mbox{$\bs{90}$-mg/$\bs{47}$-$\ts{mm}^{\bs{3}}$} VLEIBot\textsuperscript{+}, which is driven by two propulsors and fully controllable in the \textit{\mbox{two-dimensional}} ($\bs{2}$D) space. The \mbox{VLEIBot\textsuperscript{+}} can swim at speeds as high as \mbox{$\bs{16.1}\,\ts{mm} \cdot \ts{s}^{\bs{-1}}$ ($\bs{0.35}\,\ts{Bl} \cdot \ts{s}^{\bs{-1}}$)}, when driven with \mbox{heuristically-optimized} propulsors, and achieves turning rates as high as \mbox{$\bs{0.28}\,\ts{rad} \cdot \ts{s}^{\bs{-1}}$}, when tracking path references. The measured \mbox{\textit{root-mean-square}} (RMS) values of the tracking errors are as low as \mbox{$\bs{4}\,\ts{mm}$}. 
\end{abstract}

\section{Introduction} 
\label{SECTION01}
\begin{figure}[t!]
\vspace{1.2ex}
\begin{center}
\includegraphics[width=0.48\textwidth]{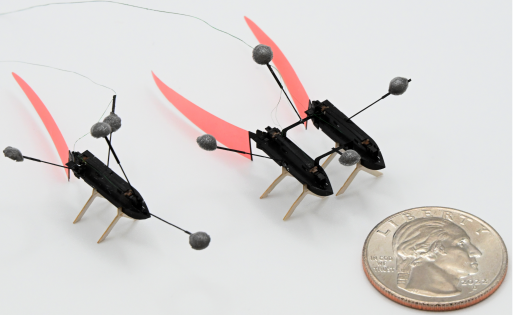}
\end{center}
\vspace{-2ex}
\caption{\textbf{Photograph of the VLEIBot and VLEIBot\textsuperscript{+}.} The VLEIBot~(left) is a \mbox{$45$-mg} swimming microrobot driven by a new bioinspired propulsor. The VLEIBot\textsuperscript{+}~(right) is a \mbox{$90$-mg} controllable microrobotic swimmer designed to achieve high maneuverability in the \mbox{\textit{two-dimensional}}~($2$D) space. \label{FIG01}}
\vspace{-2ex}
\end{figure}

The recent development of extremely light (\mbox{$1$~to~$10$\,mg}) and small (\mbox{$0.45$~to~$1.89$\,mm$^3$}) fast (up to $40\,\ts{Hz}$) actuators based on \textit{\mbox{shape-memory} alloy} (SMA) technology enabled the creation of new types of microrobots, which can operate at low voltages and with simple excitation \mbox{electronics \cite{SMALLBug_2020,SMARTI_2021,WaterStrider_2023}}. These advancements predict that the vision of creating swarms and schools of \mbox{insect-scale} robots capable of performing essential tasks for humans will become a reality soon. Here, we present two new microrobotic surface swimmers, whose propulsion mechanism is inspired by the anguilliform swimming mode at a moderate (\mbox{$\sim\hspace{-0.4ex}10^2$~to~$\sim\hspace{-0.4ex}10^4$}) \textit{Reynolds~number} ($Re$). These microrobots, shown in Fig.\,\ref{FIG01}, are: (i)~the \mbox{VLEIBot}, which weighs $45$\,mg, operates at frequencies up to $20$\,Hz, and achieves speeds as high as \mbox{$15.1\,\ts{mm} \cdot \ts{s}^{-1}$ ($0.33\, \ts{Bl} \cdot \ts{s}^{-1}$)}; and, (ii)~the \mbox{VLEIBot\textsuperscript{+}}, which weighs $90\,$\,mg, operates at frequencies up to $10$\,Hz, and achieves speeds as high as
\mbox{$16.1\,\ts{mm} \cdot \ts{s}^{-1}$ ($0.35\, \ts{Bl} \cdot \ts{s}^{-1}$)}. Additionally, because of its dual propulsion mechanism, the \mbox{VLEIBot\textsuperscript{+}} is fully controllable in the \mbox{\textit{two-dimensional}} ($2$D) space. We envision that in coordination with other flying and crawling \mbox{insect-scale} robots---advanced evolutions of those \mbox{in~\cite{SMALLBug_2020,SMARTI_2021,WaterStrider_2023,PerezJBB01,Nestor_JINT_2015,BeePlus_2019,JINT_2022,BeePlusPlus_2023,RoBeetle_2020,Fuller2017Dampers,Perez2015ModelFree,Perez2013ICAR,Duhamel2013OpticalFlow,Teoh2012IROS,Duhamel2012ICRA,Perez2011ACC,Perez2011ROBIO,Perez2013Transactions}}---teams of \mbox{VLEIBot-type} microswimmers will assist humans in aquatic search and rescue missions, inspection of vessels and marine infrastructure, hydroponic agriculture, aquaculture tasks, research of \mbox{shallow-water} reefs, and \mbox{minimally-invasive} continuous automatic monitoring of water quality in reservoirs, just to mention a few. 

The VLEIBot design exploits the surface tension of water to stay afloat and the \textit{\mbox{fluid-structure interaction}} (FSI) of a flexible thin undulating tail with water to propel itself forward. In this case, the propulsor is driven by the \mbox{$6$-mg} \textit{\mbox{high-work-density}} (HWD) \mbox{SMA-based} actuator introduced in~\cite{SMALLBug_2020} and its design was loosely inspired by the anguilliform swimming mode exhibited by European eels (\textit{Anguilla~anguilla} L.)~\cite{Palstra_EuropeanEel_2010}, \mbox{yellow-bellied} sea snakes (\textit{Pelamis~platurus})~\cite{Graham_SeaSnake_1986}, and young \textit{African clawed frog} (ACF) tadpoles (\textit{Xenopus~laevis})~\cite{Roberts_XenopusLaevis_2000}. Anguilliform swimming is believed to be generated by \mbox{sub-ambient} pressure zones in the troughs of the swimmer's body waves and, for this reason, highly efficient regarding cost of transport. This is an FSI phenomenon observed at moderate $Re$s (\mbox{$\sim\hspace{-0.4ex}10^2$ to $\sim\hspace{-0.4ex}10^4$})~\mbox{\cite{Tack_Anguilliform_2021,Liu_TadpoleCFD_1997}}. Given that the $Re$s associated with the motion of the VLEIBot's propulsor are in the moderate range (\mbox{$\sim\hspace{-0.4ex}10^{-1}$ to $\sim\hspace{-0.4ex}10^{3}$}), we hypothesize that the same hydrodynamic effect occurs in this case and we considered it during the design process. In the past, researchers have distributively embedded SMA wires in \mbox{semi-rigid} bodies to generate undulation~\cite{Fukuda_DistributedSMA_1990,Garner_SMABiomimetic_2000,Rossi_BendingSMA_2011,Cho_Body_Caudal_2008,WANG_MicroFish_2008,Shi_Jellyfish_2010}; however, to our best knowledge, the VLEIBot's design is the first with onboard actuation that removes the need for complex propulsor configurations and fully exploits FSI phenomena to generate a wave that propagates along the swimmer's tail, thus producing locomotion. This is the key design element that enables \mbox{anguilliform-inspired} swimming at the \mbox{mm-to-cm} scale for the first time.

The most common actuation methods used to drive swimming microrobots are based on electromagnetic~\cite{Byun_Helmholtz_2012,Zhao_OrigamiSwimmer_2022,Chen_MicroSwimmers_2023,Tan_MagneticSwimmer_2023,Seyed_Hydrogel_2011,Fusco_IntegratedMicrorobot_2014,Li_ControlledDrug_2009,Temel_ConfinedSwimming_2015,Palagi_SoftMagSwimmer_2011,Pak_HighSpeedNanowire_2011,DreyfusMicroArtSwimmer_2005,GaoFlexibleNanowireMotor_2010,Xu2013HelicalMicroswimmers, Ghosh2009ControlledPropulsion, Sing2009CollidalWalkers, Zhang2010RotatingNickleNanowires, Liu2010WirelessSwimmingMicrorobot, Tierno2008ColloidalMicroswimmers, Floyd2008UntetheredMagneticallyActuated}, piezoelectric~\cite{Zhao2021PZTFrog, Li2023PZT, song2007PZTWaterStrider, Chen2018HybridTerrestrialAquaticMicrorobot, Deng2005OscillatingFinPropulsor, Ming2009PiezoelectricFiberComposite, Fukuda1994MicroMobileRobots}, and \mbox{ionic-polymer}~\cite{Kim_2005, Kamamichi2006SwimmingSnake, Guo2003FishLikeMicrorobot} technologies. While electromagnetic actuation has enabled the functionality of very small and fast swimmers, some of which can operate inside the human body, this approach does not represent a path to full autonomy at the \mbox{mm-to-cm} scale because it requires the generation of large external magnetic fields in laboratory environments. The main advantage of piezoelectric actuation is its wide frequency bandwidth; however, it requires very high excitation voltages (\mbox{$\sim\hspace{-0.4ex}300\,\ts{V}$}) and, therefore, complex excitation electronics. Wide actuation bandwidths are required for creating \mbox{insect-scale} \mbox{flapping-wing} flyers, but the observed undulation frequencies in anguilliform swimmers do not overpass $10\,\ts{Hz}$~\cite{Tack_Anguilliform_2021}. \mbox{Ionic-polymer} actuation provides unique new capabilities for soft robotic systems; however, its relatively low force outputs significantly limit its applicability to the development of aquatic propulsion systems. In contrast with other actuators for microrobotics, the actuator that drives the VLEIBot is not only light ($6\,\ts{mg}$) and fast (up to $30\,\ts{Hz}$), but also can generate large forces (\mbox{$\sim\hspace{-0.4ex}84\,\ts{mN}$}) using very low voltages \mbox{($5$ to $25\,\ts{V}$)} and simple excitation electronics. To heuristically optimize the hydrodynamic design and operation parameters of the VLEIBot, we used a Vicon \mbox{motion-capture} system to investigate the relationship between the robot's average speed and tail geometry through a series of characterization experiments. We discovered that, among the cases tested, the fastest speed of \mbox{$15.1\,\ts{mm} \cdot \ts{s}^{-1}$ ($0.33\,\ts{Bl} \cdot \ts{s}^{\bs{-1}}$)} is attained using a parabolic tail with a constant aspect ratio of $0.41$ and length of $26\,\ts{mm}$, undulating at a frequency of $5\,\ts{Hz}$.

Simple swimming experiments showed that, due to asymmetrical force production of the \mbox{SMA-based} exciting actuator and high sensitivity to external disturbances, the VLEIBot exhibits low maneuverability. To address this performance challenge, we developed the \mbox{$90$-mg} VLEIBot\textsuperscript{+}, a modular design that achieves full $2$D controllability by using two \mbox{$6$-mg} SMA actuators that independently drive two \mbox{anguilliform-inspired} propulsors. From a control perspective, this approach follows the methodology first introduced in~\cite{SMARTI_2021}. In the case presented here, we achieved high turning capabilities. Using a Vicon \mbox{motion-capture} system, to evaluate and demonstrate the locomotion performance that the VLEIBot\textsuperscript{+} can achieve, we performed several swimming \mbox{open-loop} and \mbox{closed-loop} control experiments. In addition to swimming at speeds on the order of \mbox{$16\,\ts{mm} \cdot \ts{s}^{-1}$ ($0.35\,\ts{Bl} \cdot \ts{s}^{\bs{-1}}$)}, the VLEIBot\textsuperscript{+} can turn at rates as high as $0.28\,\ts{rad} \cdot \ts{s}^{-1}$. The measured \mbox{\textit{root-mean-square}} (RMS) \mbox{tracking-error} values are as low as $3.94\,\ts{mm}$ when operating at the optimal frequency of $5\,\ts{Hz}$. These results are highly significant because it is well known that as the $Re$ decreases, swimming becomes increasingly challenging by the traditional methods used at the human scale ($Re \approx 10^4$)~\cite{PurcellAJP_1977}. The VLEIBot and VLEIBot\textsuperscript{+} highlight the possibility of creating \mbox{mm-to-cm--scale} autonomous swimming microrobots that can work together in schools (teams) to perform complex tasks useful to humans. 

The rest of the paper is organized as follows. Section\,\ref{SECTION02} presents the design and fabrication of the \mbox{$45$-mg} VLEIBot. Section\,\ref{SECTION03} describes the experimental setup used in the swimming experiments, and discusses the characterization and performance evaluation of two sets of bioinspired soft propulsors. Section\,\ref{SECTION04} discusses the development of the VLEIBot\textsuperscript{+} and presents \mbox{closed-loop} swimming experiments. Finally, Section\,\ref{SECTION05} states some conclusions regarding the presented research.
\begin{figure}[t!]
\vspace{1.2ex}
\begin{center}
\includegraphics[width=0.48\textwidth]{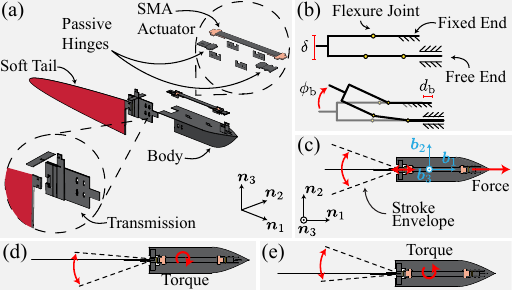}
\end{center}
\vspace{-2ex}
\caption{\textbf{Robotic design and functionality.} \textbf{(a)}~Components and assembly of the VLEIBot. The VLEIBot is composed of four main components: (i)~a rigid body that supports the weight of the robot on water due to surface tension; (ii)~a \mbox{$6$-mg} HWD \mbox{SMA-based} actuator with passive hinges installed at its two distal ends; (iii)~a planar \mbox{four-bar} transmission mechanism that maps the output displacement generated by the \mbox{SMA-based} actuator into the angular oscillation that excites the undulating tail of the swimmer; and, (iv)~an \mbox{anguilliform-inspired} soft tail made of fluoropolymer film. The triplet \mbox{$\left\{ \bs{n}_1,\bs{n}_2,\bs{n}_3 \right\}$} denotes the inertial frame used to describe the kinematics of the system. \textbf{(b)}~Transmission functionality. The planar \mbox{four-bar} mechanism can be adjusted to have a constant bias angle, $\phi_\ts{b}$, by shifting the installation point of the fixed end by a distance $d_\ts{b}$. \textbf{(c)}~Forward locomotion. The VLEIBot is designed to function with a symmetric stroke envelope (sweeping area). The \mbox{SMA-based} actuator produces the periodic displacement output that is mapped by the transmission into the large stroke angles required to undulate the robot's tail. The triplet \mbox{$\left\{ \bs{b}_1,\bs{b}_2,\bs{b}_3 \right\}$} denotes the \mbox{body-fixed} frame used to describe the kinematics of the system. \textbf{(d)}~Right turn. Theoretically, a \mbox{right-biased} undulation produces a negative torque with respect to the body frame and, as a consequence, a right turn. \textbf{(e)}~Left turn. Theoretically, a \mbox{left-biased} undulation produces a positive torque with respect to the body frame and, as a consequence, a left turn.
\label{FIG02}}
\vspace{-2ex}
\end{figure}

\section{Design and Fabrication} 
\label{SECTION02}
As shown in Fig.\,\ref{FIG02}(a), the VLEIBot has four main components: \mbox{(i)~a} rigid body that supports the weight of the entire robot on water due to surface tension; \mbox{(ii)~a} \mbox{$6$-mg} \mbox{SMA-based} actuator, which is installed in the robot with passive hinges attached to its two distal ends; \mbox{(iii)~a} planar \mbox{four-bar} transmission; and, \mbox{(iv)~a} soft tail heuristically designed to generate \mbox{anguilliform-inspired} locomotion. All composing parts were designed and fabricated using the methods described in~\cite{BeePlus_2019,RoBeetle_2020,BeePlusPlus_2023,JINT_2022} and references therein. The body of the robot is made of \textit{carbon fiber} (CF); the actuator is made of CF, FR4, and SMA wires, according to the configuration in~\cite{SMALLBug_2020,SMARTI_2021}; the transmission and hinges are made of CF and Kapton; and, the soft tail is made of \mbox{$25$-{\textmu}m-thick} fluoropolymer film (AirTech~A$4000$R$14417$). The basic configuration of the body is simply the perpendicular intersection of two flat CF pieces with significant areas relative to their volumes. One piece interacts horizontally with the water and, because of a relatively high surface tension, maintains the robot afloat; the other piece functions as a keel, thus providing stability and high lateral drag compared to that acting on the tail. The \mbox{$6$-mg} HWD \mbox{SMA-based} actuator excites the \mbox{four-bar} transmission; the transmission excites the flexible tail, which periodically undulates during swimming to produce thrust. To select the geometrical parameters that maximize the transmission ratio, we used the nonlinear relationship discussed in~\cite{Finio2011SystemID}. Also, we incorporate interchangeability and enforce alignment in the tail installation process by using a keyway system integrated into the transmission.

As discussed in~\cite{SMALLBug_2020}, actuators of the type employed to drive the VLEIBot accumulate a \mbox{steady-state} bias during operation at \textit{\mbox{pulse-width} modulation} (PWM) frequencies higher than $1\,\ts{Hz}$ because of the limited time available for the composing SMA wires to cool down within an actuation cycle. To address this issue, the transmission is tuned and installed with a fixed bias angle, $\phi_{\ts{b}}$, as depicted in Fig.\,\ref{FIG02}(b). This bias angle is created by simply displacing the fixed end of the transmission, as graphically defined in Fig.\,\ref{FIG02}(b), by a distance $d_\ts{b}$; namely, $\phi_{\ts{b}}=T\cdot d_{\ts{b}}$, where $T$ is the transmission ratio. For small displacements, the transmission ratio can be estimated as \mbox{$T = \delta^{-1}$}~\cite{Finio2011SystemID}, where $\delta$ is the offset between the fixed and free ends of the mechanism, as graphically defined in Fig.\,\ref{FIG02}(b). In the case of \mbox{PWM-based} Joule heating, because of the $\phi_{\ts{b}}$ adjustment, when the actuator contracts during an actuation cycle, the direction of bending is opposite to that of $\phi_{\ts{b}}$. This effect allows us to generate a symmetric stroke envelope (sweeping area) with respect to the \mbox{body-fixed} $\bs{b}_1$-$\bs{b}_3$ plane, for any given operational frequency, as depicted in Fig.\,\ref{FIG02}(c). Unfortunately, different operational frequencies require different $\phi_{\ts{b}}$ adjustments. Nonetheless, test results indicate that the VLEIBot can achieve straight forward locomotion in open loop after its transmission is tuned for a predefined operational condition. The tuning procedure is very simple. First, we install the transmission with a bias angle of approximately $0.3$\,rad; then, we excite the VLEIBot with a signal that produces significant forward locomotion, selected through the tests presented and discussed in Section\,\ref{SECTION03}; last, we visually determine the magnitude and direction of $\phi_\ts{b}$ required to adjust the transmission. We repeat this process, if necessary, until we observe a symmetric stroke envelope. 

By generating a symmetric stroke envelope, we ensure that the total hydrodynamic force produced by the undulating tail has a lateral average of zero in steady state and, therefore, its resulting average direction is ideally aligned with the \mbox{body-fixed} $\bs{b}_1$ axis. However, due to external disturbances, such as the pull of the tether wire and wind gusts, the VLEIBot often deviates from a given straight path in \mbox{open loop} operation. This observation highlights the need for feedback control, even after the stroke envelope has been tuned. Theoretically, basic control strategies can be implemented by simply modulating the PWM \mbox{actuator-exciting} signal and thus create an asymmetrical tail undulation with respect to the \mbox{body-fixed} $\bs{b}_1$ axis (see Fig.\,\ref{FIG02}(c)). As illustrated in \mbox{Figs.\,\ref{FIG02}(d)~and~(e)}, theoretically, an asymmetrical stroke envelope produces a torque that rotates the robot during swimming, thus endowing it with turning capabilities and controllability in $2$D. Unfortunately, empirical evidence indicates that this turning ability of the robot is highly sensitive to slight miscalibration of the stroke envelope and external disturbances. Fortunately, as discussed in Section\,\ref{SECTION04}, this steerability issue can be solved with a modular design that puts together two VLEIBot platforms to create the VLEIBot\textsuperscript{+}.
\begin{figure}[t!]  
\vspace{1.2ex}
\begin{center}
\includegraphics[width=0.48\textwidth]{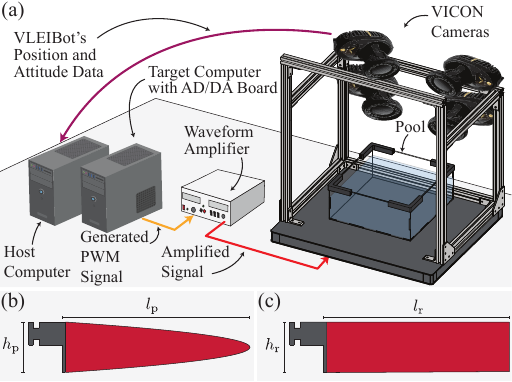}
\end{center}    
\vspace{-2ex}
\caption{\textbf{Experimental setup used in the \mbox{tail-characterization} swimming tests.} \textbf{(a)}~Experimental setup. A Simulink \mbox{Real-Time} \mbox{host-target} system and an \mbox{AD/DA} board (National Instruments PCI-$6229$) are used to generate the PWM signal that excites the robot. This signal is then filtered through a waveform amplifier (Accel Instruments TS$250$-$02$) to provide the power necessary to periodically Joule heat the \mbox{SMA-based} actuator that drives the VLEIBot. During the swimming tests, the VLEIBot is placed in a pool filled with water and a \mbox{four-VK16-camera} Vicon \mbox{motion-capture} system is used to measure the instantaneous position and orientation of the robot at a rate of $250\,\ts{Hz}$. \textbf{(b)}~Parameters of the parabolic tails with constant aspect ratio. For the \mbox{tail-characterization} experiments, we kept the aspect ratio constant at $\AR_{\ts{p}} = 0.41$ and varied the tail length, $l_{\ts{p}}$, in increments of $2\,\ts{mm}$ over the range $\left[0:28\right]\,\ts{mm}$. The height $h_{\ts{p}}$ directly depends on the length $l_{\ts{p}}$. \textbf{(c)}~Parameters of the rectangular tails with constant height. For the \mbox{tail-characterization} experiments, we kept the height, $h_\ts{r}$, constant at $4\,\ts{mm}$ and varied the length, $l_\ts{r}$, in increments of $5\,\ts{mm}$ over the range $[0:50]\,\ts{mm}$. \label{FIG03}}
\vspace{-2ex}
\end{figure}

\section{Tail Characterization}
\label{SECTION03}
\subsection{Experimental Setup and Swimming Tests}
\label{SECTION03A}
An illustration of the experimental setup used in the swimming tests is shown in~Fig.\,\ref{FIG03}(a). As seen here, to generate the PWM signal that excites the robot, and for signal processing and control, we use a Simulink \mbox{Real-Time} \mbox{host-target} system and a National Instruments \mbox{PCI-$6229$} \mbox{\textit{analog-digital/digital-analog}}~(AD/DA) board. After the PWM signal is generated, this is filtered through a waveform amplifier (Accel Instruments TS$250$-$02$) to supply the power necessary to cyclically Joule heat the SMA material of the robot's actuator. During the experiments, a \mbox{VLEIBot} prototype is placed in a pool filled with water while a \mbox{four-VK16-camera} Vicon \mbox{motion-capture} system measures and records its instantaneous position, $\bs{r}$, and attitude at a frequency of $250\,\ts{Hz}$. We estimate the velocity, $\bs{v}$, and angular velocity, $\bs{\omega}$, of the robot using a \mbox{discrete-time} derivative algorithm and, then, these data are further processed using a \mbox{low-pass} \textit{finite impulse response} (FIR) filter with order $10^3$ and cutoff frequency of $40\,\ts{Hz}$. For consistency with the Vicon system, the \mbox{host-target} system samples, processes, generates, and records signals at $250\,\ts{Hz}$. 

To characterize the effects of tail geometry and size on the swimming performance of the VLEIBot, we experimentally investigated two basic tail geometries: (i)~a \mbox{low-aspect-ratio} \mbox{parabolic-shaped} tail, loosely inspired by the tails of the European eel and young ACF tadpole; and, (ii)~a thin \mbox{rectangular-shaped} tail, loosely inspired by the tail of the \mbox{yellow-bellied} sea snake. During the characterization process, we tested several aspect ratios of the two basic geometries selected, over a range of excitation frequencies. Then, using the measured data, we heuristically determined the \mbox{best-performing} tail for the VLEIBot regarding speed of locomotion. In the \mbox{tail-characterization} experiments and analyses, we estimate the Reynolds number as 
\vspace{-0.6ex}
\begin{align}
Re = \frac{l v \rho}{\mu},
\label{EQN01}
\vspace{-0.7ex}
\end{align}
in which $l$ and $v$ are the length (including the tail) and speed of the swimmer, respectively; and, $\rho$ and $\mu$ are the density and dynamic viscosity of water at $20\,^{\circ}\textrm{C}$, respectively. Considering all achieved speeds and tested tails, the $Re$ is inside the range $\left[10^{-1}: 10^3 \right]$. As a reference, the $Re$ for the swimming of the young ACF tadpole is on the order of 250~\cite{Roberts_XenopusLaevis_2000}. Also, similarly to the VLEIBot, the young ACF tadpole has a wide rigid head and a thin \mbox{highly-flexible} tail. Furthermore, the lack of osseous or cartilaginous support in their tails~\cite{Liu_TadpoleCFD_1997}, prompt us to speculate that the young ACF tadpoles might take advantage of FSI phenomena to locomote.

\begin{figure}
\vspace{1.2ex}
\begin{center}
\includegraphics[width=0.48\textwidth]{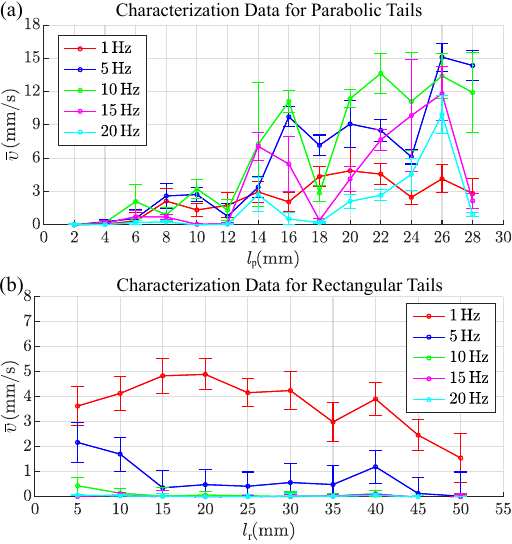}
\end{center}
\vspace{-2ex}
\caption{\textbf{\mbox{Tail-characterization} data.} \textbf{(a)}~Characterization of parabolic tails with constant aspect ratio. During the swimming tests, the aspect ratio was kept constant at \mbox{$\AR_{\ts{p}} = 0.41$}, while we investigated the relationship between tail length and average forward speed across frequencies of $1$, $5$, $10$, $15$, and $20$\,Hz. We see an increasing trend in average speed as tail length increases. At the tail length of $26\,\ts{mm}$ and frequency of $5\,\ts{Hz}$, we measured the maximum average speed of \mbox{$15.1\,\ts{mm} \cdot \ts{s}^{-1}$ ($0.33\,\ts{Bl} \cdot \ts{s}^{-1}$)}. When the tail length, $l_{\ts{p}}$, was increased further, the weight of the tail tended to drag the robot under the surface of the water. \textbf{(b)}~Characterization of rectangular tails with constant height. During the swimming tests, the tail height was kept constant at \mbox{$h_{\ts{r}} = 4\,\ts{mm}$}, while we investigated the relationship between tail length and speed. We varied the tail length in increments of \mbox{$5\,\ts{mm}$} over the range \mbox{$\left[5:50\right]\,\ts{mm}$}. It can be observed that \mbox{high-frequency} actuation is not able to produce thrust with any of the investigated tails, and \mbox{$1\,\ts{Hz}$} is the best actuation frequency for all the tested profiles. At the length of \mbox{$20\,\ts{mm}$}, we see the best swimming performance of \mbox{$4.7\,\ts{mm} \cdot \ts{s}^{-1}$} \mbox{($0.12\,\ts{Bl} \cdot \ts{s}^{-1}$)}.
\label{FIG04}}
\vspace{-2ex}
\end{figure}

The similarities between the VLEIBot and young ACF tadpoles are the main reasons to use the anatomy of these animals as a model for tail design. Specifically, we selected a constant aspect ratio, specified as $\AR_{\ts{p}} = h_{\ts{p}}^2 \cdot A_{\ts{p}}^{-1}$, where $h_{\ts{p}}$ and $A_{\ts{p}}$ are respectively the height and area of the tail, both graphically defined in Fig.\,\ref{FIG03}(b). Assuming a parabolic shape, we determined that the aspect ratio is approximately $0.41$ for the tails of young ACF tadpoles. By keeping the aspect ratio of each tested tail fixed, its height becomes a function of its length, $l_{\ts{p}}$, which is defined in~Fig.\,\ref{FIG03}(b). With this in mind, for the constant aspect ratio $\AR_{\ts{p}} = 0.41$, we made and tested $14$ different tails in the set \mbox{$\left[2:2:28\right]$\,mm}. To design the tested rectangular tails, loosely inspired by the tail geometry of \mbox{yellow-bellied} sea snakes, we simply selected the constant height \mbox{$h_\ts{r}=4$\,mm}, as defined in Fig.\,\ref{FIG03}(c), and varied the length, $l_\ts{r}$, as illustrated in Fig.\,\ref{FIG03}(c), over the range $\left[5:50\right]$\,mm in increments of $5\,\ts{mm}$. To evaluate the performance associated with the tested tails, we measured the average forward speed of the robot across all the considered tail lengths and shapes. For each tested pair of exciting frequency and tail shape, we performed $10$ swimming tests. The tested frequencies are \mbox{$1$, $5$, $10$, $15$, and $20$\,Hz} with a heuristically chosen duty cycle---defined as the percentage of the PWM period for which the signal is at its \textit{on} voltage---of $8\,\%$, for the excitation of $1\,\ts{Hz}$, and of $10\,\%$, for the other tested frequencies. An \textit{on} voltage of $22\,\ts{V}$ was selected such that the actuator is provided with $250\,\ts{mA}$ of current; higher current values were observed to burn the thin $53$\,AWG tether wires that power the robot. For analysis, the average speed of each swimming experiment was computed from $20\,\ts{s}$ of data. Then, we found the mean and \textit{empirical standard deviation}~(ESD) of the $10$ tests corresponding to each pair of PWM frequency and tail shape. These data are summarized in Fig.\,\ref{FIG04} and discussed next.
\begin{figure*}[t!]
\vspace{1.2ex}
\begin{center}
\includegraphics[width=0.99\textwidth]{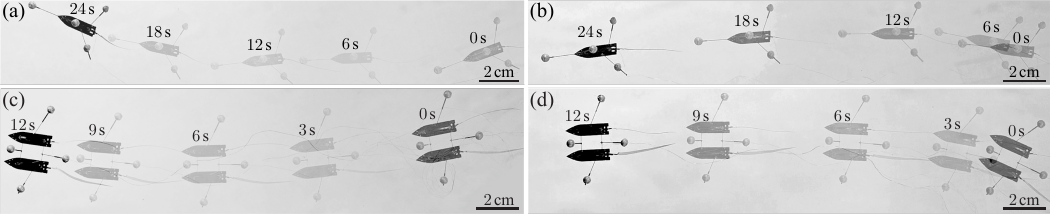}
\end{center}
\vspace{-2ex}
\caption{\textbf{Locomotion experiments performed using optimized parabolic $\bs{26}$-mm tail with constant aspect ratio of $\bs{0.41}$.} \textbf{(a)}~Photographic composite from video footage showing the VLEIBot swimming at $1\,\ts{Hz}$. The duty cycle of the driving PWM signal is $8\,\%$. \textbf{(b)}~Photographic composite from video footage showing the VLEIBot swimming at $5\,\ts{Hz}$. The duty cycle of the driving PWM signal is $12\,\%$. \textbf{(c)}~Photographic composite from video footage showing the VLEIBot\textsuperscript{+} swimming at $1$\,Hz. The duty cycles of the driving PWM signals are $8\,\%$. \textbf{(d)}~Photographic composite from video footage showing the VLEIBot\textsuperscript{+} swimming at $5\,\ts{Hz}$. The duty cycles of the driving PWM signals are $12\,\%$. Video footage of these experiments can be viewed in the accompanying supplementary movie. \label{FIG05}}
\vspace{-2ex}
\end{figure*}

\subsection{Results and Discussion}
\label{SECTION03B}
The experimental data corresponding to the parabolic tails with a constant aspect ratio \mbox{$\AR_{\ts{p}} = 0.41$} are shown in Fig.\,\ref{FIG04}(a). Here, we can clearly see a trend; over the range $\left[2:26\right]\,\ts{mm}$, as the tail length, $l_{\ts{p}}$, increases, the average speed, $\bar{v}$, tends to increase. After passing a length of $26\,\ts{mm}$, $\bar{v}$ tends to decrease because of the relatively high weight of the tail compared to that of the robot. With a frequency of $5\,\ts{Hz}$ and \mbox{$l_{\ts{p}} = 26\,\ts{mm}$}, we obtained the best swimming performance in terms of speed, \mbox{$15.1\,\ts{mm} \cdot \ts{s}^{-1}$ ($0.33\,\ts{Bl} \cdot \ts{s}^{-1}$)}. The experimental data corresponding to the rectangular tails with constant height \mbox{$h_{\ts{r}} = 4\,\ts{mm}$} are shown in Fig.\,\ref{FIG04}(b). In these tests, for all tail lengths, the best frequency is $1\,\ts{Hz}$; for this frequency, the best average speed is \mbox{$4.7\,\ts{mm} \cdot \ts{s}^{-1}$ ($0.12\,\ts{Bl} \cdot \ts{s}^{-1}$)}, corresponding to a tail length of $20\,\ts{mm}$. At frequencies higher than $5\,\ts{Hz}$, essentially, no locomotion can be produced with the tested rectangular tails. Overall, the swimming performances achieved with the parabolic tails are significantly better than those obtained with the rectangular tails. Through this heuristic process, we selected the parabolic \mbox{$26$-mm} tail for the VLEIBot prototype. Photographic composites of frames taken at intervals of $6\,\ts{s}$ over $24\,\ts{s}$, with the \textit{optimized} VLEIBot swimming at $1$ and $5\,\ts{Hz}$ are shown in \mbox{Figs.\,\ref{FIG05}(a)~and~(b)}, respectively. Video footage of these two experiments are shown in the accompanying supplementary movie. 

\section{The VLEIBot\textsuperscript{+}: A Control Approach}
\label{SECTION04}
\subsection{Controllable Design}
\label{SECTION04A}
\begin{figure*}[ht!]
\vspace{1.2ex}
\begin{center}   
\includegraphics[width=0.99\textwidth]{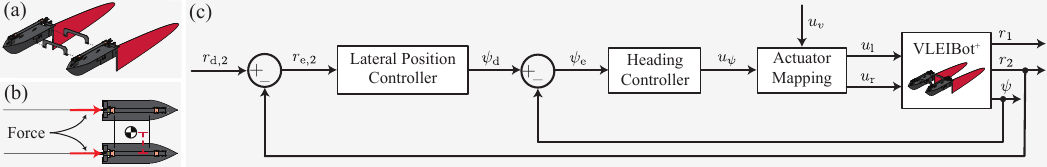}
\end{center}
\vspace{-2ex}
\caption{\textbf{The VLEIBot\textsuperscript{+} and block diagram of the control scheme.} \textbf{(a)}~Assembly of two VLEIBots in parallel to create the VLEIBot\textsuperscript{+}. \textbf{(b)}~Turning functionality of the VLEIBot\textsuperscript{+}. The separation between the two propulsors of the VLEIBot\textsuperscript{+} enables a dependable production of turning torque by introducing a net force differential. \textbf{(c)}~Architecture used to control the VLEIBot\textsuperscript{+} during swimming. The lateral position controller receives as input the lateral position error, \mbox{$r_{\ts{e},2} = r_{\ts{d},2} - r_2$}, and computes the desired yaw angle, $\psi_{\ts{d}}$. Next, the heading controller receives as input the \mbox{yaw-angle} error, \mbox{$\psi_\ts{e} = \psi_\ts{d} - \psi$}, and computes the heading control signal, $u_{\psi}$. Last, the actuator mapping receives as inputs $u_{\psi}$ and the constant $u_v$, and maps them into the duty cycles $u_\ts{l}$ and $u_\ts{r}$, which are used to define the PWM signals that excite the left and right \mbox{SMA-based} actuators of the robot, respectively.
\label{FIG06}}
\end{figure*}

As explained in previous sections, after multiple \mbox{open-loop} swimming experiments, we determined that the VLEIBot design lacks the steerability required to execute \mbox{high-speed} turning maneuvers. To address this issue, we developed the \mbox{$90$-mg} \mbox{dual-propulsor} VLEIBot\textsuperscript{+}, which is fast, highly maneuverable, and fully controllable in the $2$D space. Photographic composites of frames taken at intervals of $3\,\ts{s}$ over $12\,\ts{s}$, with a VLEIBot\textsuperscript{+} prototype swimming at $1$ and $5\,\ts{Hz}$, are shown in \mbox{Figs.\,\ref{FIG05}(c)~and~(d)}, respectively. Video footage of these two experiments are shown in the accompanying supplementary movie. A VLEIBot\textsuperscript{+} is built by connecting two VLEIBot prototypes in the parallel and symmetrical configuration depicted in~Fig.\,\ref{FIG06}(a), using two CF beams. The separation between the two composing bodies is chosen to be $6\,\ts{mm}$ in order to enable the cancelation, due to symmetry, of undesired torques produced by the undulating tails during forward locomotion; this notion is illustrated in Fig.\,\ref{FIG06}(b). The swimming tests corresponding to \mbox{Figs.\,\ref{FIG05}(c)~and~(d)} compellingly demonstrate an \textit{almost} perfect cancelation of the opposing torques generated by the two sides of the tested prototype. Due to this effect, VLEIBot\textsuperscript{+} prototypes are fully controllable in the $2$D space.

\subsection{Control Strategy}
\label{SECTION04B}
To enable the VLEIBot\textsuperscript{+} to execute turning maneuvers and track references in the $2$D space, we implemented the control scheme in Fig.\,\ref{FIG06}(c), which is almost identical to that in~\cite{SMARTI_2021}. Here, the pair $\left\{r_1,r_2\right\}$ (first two components of $\bs{r}$) is the instantaneous position of the robot's \mbox{body-fixed} frame in the $2$D space relative to the inertial frame, and $\psi$ is the instantaneous yaw angle, measured with the Vicon system. In this scheme, the lateral position controller computes the desired heading \mbox{according to}
\vspace{-0.6ex}
\begin{align}
\psi_\ts{d} = k_{\ts{p},2} r_{\ts{e},2} + k_{\ts{i},2}\int_0^t r_{\ts{e},2}\,d\tau,
\label{EQN02}
\vspace{-0.7ex}
\end{align}
where $k_{\ts{p},2}$ and $k_{\ts{i},2}$ are proportional and integral controller gains; and, \mbox{$r_{\ts{e},2} = r_{\ts{d},2} - r_{2}$} is the \mbox{lateral-position} error computed as the difference between the desired and measured lateral positions, $r_{\ts{d},2}$ and $r_2$, respectively. The heading controller computes the control signal \mbox{$u_{\psi} = k_{\ts{p},\psi} \psi_\ts{e}$}, where $k_{\ts{p},\psi}$ is a proportional controller gain and \mbox{$\psi_\ts{e} = \psi_\ts{d} - \psi$} is the \mbox{yaw-angle} error computed as the difference between the desired and current yaw angles, $\psi_\ts{d}$ and $\psi$, respectively. Last, the actuator mapping receives $u_{\psi}$ and a constant $u_v$, and maps them into the \mbox{duty-cycle} values $u_\ts{l}$ and $u_\ts{r}$ used to define the PWM signals that respectively drive the left and right actuators of the robot, according to
\begin{align}
\left[
    \begin{array}{c}
    u_\ts{l}\\
    \vspace{-2ex}
    \\
    u_\ts{r}\\
    \end{array} 
    \right]= \hspace{-0.6ex}
\left[
    \begin{array}{cc}
    k_\ts{l} & -k_\ts{l}\\
    \vspace{-2ex}
    \\
    k_\ts{r} & k_\ts{r}\\
    \end{array}
    \right]\hspace{-0.3ex}
    \left[
    \begin{array}{c}
    u_v\\
    \vspace{-2ex}
    \\
    u_\psi\\
    \end{array}
    \right],
\label{EQN03}
\end{align}
where $k_\ts{l}$ and $k_\ts{r}$ are tunable mapping constants. 
\begin{figure*}[t!]
\vspace{1.2ex}
\begin{center}
\includegraphics[width=0.99\textwidth]{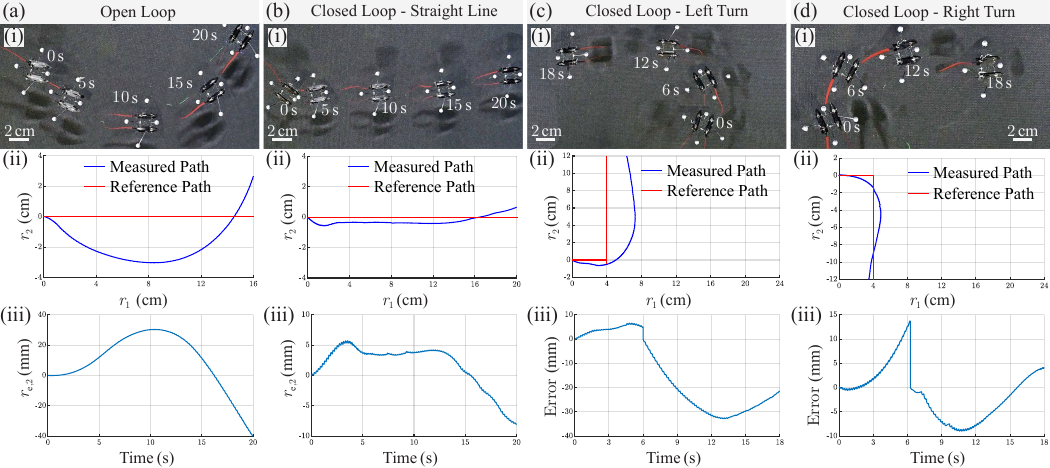}
\end{center}
\vspace{-2ex}
\caption{\textbf{VLEIBot\textsuperscript{+} locomotion experiments.} \textbf{(a)}~$20$-s \mbox{open-loop} swimming experiment at $5$\,Hz. (i)~Photographic composite of frames taken at intervals of $5\,\ts{s}$ from video footage of the VLEIBot\textsuperscript{+} swimming forward in open loop. Both actuators of the robot were excited using a $5$-Hz PWM signal with a duty cycle of $12\,\%$. (ii)~Measured position of the VLEIBot\textsuperscript{+} over the course of the \mbox{open-loop} experiment. (iii)~Lateral error of the VLEIBot\textsuperscript{+} over the course of the experiment. We measured an \mbox{RMS-error} value of $20.4\,\ts{mm}$. \textbf{(b)}~$20$-s \mbox{closed-loop} swimming experiment at $5\,\ts{Hz}$. (i)~Photographic composite of frames taken at intervals of $5\,\ts{s}$ from video footage of the VLEIBot\textsuperscript{+} swimming forward in closed loop. (ii)~Measured position of the VLEIBot\textsuperscript{+} over the course of the \mbox{closed-loop} experiment. (iii)~Lateral error of the VLEIBot\textsuperscript{+} over the course of the experiment. We measured an \mbox{RMS-error} value of $3.94\,\ts{mm}$. \textbf{(c)}~$18$-s \mbox{closed-loop} \mbox{left-turn} experiment at $5\,\ts{Hz}$. (i)~Photographic composite of frames taken at intervals of $6\,\ts{s}$ from video footage of the VLEIBot\textsuperscript{+} swimming in closed loop. (ii)~Measured position of the VLEIBot\textsuperscript{+} over the course of the \mbox{closed-loop} experiment. We measure a maximum angular turning rate of $0.25\,\ts{rad} \cdot \ts{s}^{-1}$. (iii)~Lateral error of the VLEIBot\textsuperscript{+} over the course of the experiment. We measured an \mbox{RMS-error} value of $20.8\,\ts{mm}$. \textbf{(d)}~$18$-s \mbox{closed-loop} right-turn experiment at $5\,\ts{Hz}$. (i)~Photographic composite of frames taken at intervals of $6\,\ts{s}$ from video footage of the VLEIBot\textsuperscript{+} swimming in closed loop. (ii)~Measured position of the VLEIBot\textsuperscript{+} over the course of the \mbox{closed-loop} experiment we measure a maximum angular turning rate of $0.28\,\ts{rad} \cdot \ts{s}^{-1}$. (iii)~Lateral error of the VLEIBot\textsuperscript{+} over the course of the experiment. We measured an \mbox{RMS-error} value of $5.45\,\ts{mm}$. Video footage of the experiments can be viewed in the accompanying supplementary movie.\label{FIG07}}
\vspace{-2ex}
\end{figure*}

\subsection{Swimming Experiments}
\label{SECTION04C}
To evaluate and demonstrate the swimming capabilities of the VLEIBot\textsuperscript{+}, we used the same setup depicted in Fig.\,\ref{FIG03}(a) and described in Section\,\ref{SECTION03A}. Fig.\,\ref{FIG07}(a) summarizes the experimental data of a VLEIBot\textsuperscript{+} prototype swimming for $20\,\ts{s}$ at $5\,\ts{Hz}$ in open loop, after being tuned to nominally follow a straight path. In this test, the two robot's actuators were excited using PWM signals with an \textit{on} voltage of $22\,\ts{V}$ and duty cycle of $12\,\%$. Fig.\,\ref{FIG07}(a.i) shows a photographic composite of frames taken at intervals of $5\,\ts{s}$, of the swimming test. And, \mbox{Figs.\,\ref{FIG07}(a.ii)~and~(a.iii)} show the robot's $2$D position and \mbox{lateral-position} error during the test. The maximum average forward speed, computed over a $1$\,second window, is \mbox{$13.6\,\ts{mm} \cdot \ts{s}^{-1}$ ($0.30\,\ts{Bl} \cdot \ts{s}^{-1}$)}. For this experiment, the RMS value of the \mbox{lateral-position} error is $20.4\,\ts{mm}$. Fig.\,\ref{FIG07}(b) summarizes the experimental data of a VLEIBot\textsuperscript{+} prototype swimming for $20\,\ts{s}$ at $5\,\ts{Hz}$ in closed loop, following a \mbox{straight-line} reference and driven by the controller described in Section\,\ref{SECTION04B}. In this case, \mbox{$k_{\ts{p,2}} = 2\,\ts{rad} \cdot \ts{m}^{-1}$}, \mbox{$k_{\ts{i,2}} = 0.1\,\ts{rad} \cdot \ts{m}^{-1}$}, \mbox{$k_{\psi} = 0.2\,\ts{rad}^{-1}$}, \mbox{$u_v = 0.11$} (per unit), \mbox{$k_{\ts{l}}=1$}, and \mbox{$k_{\ts{r}} = 1$}; the \textit{on} PWM voltage was set at $22\,\ts{V}$. Fig.\,\ref{FIG07}(b.i) shows a photographic composite of frames taken at intervals of $5\,\ts{s}$, of the swimming test. And, \mbox{Figs.\,\ref{FIG07}(b.ii)~and~(b.iii)} show the robot's position in the $2$D space and \mbox{lateral-position} error during the test. The maximum average forward speed is $14.8\,\ts{mm} \cdot \ts{s}^{-1}$ ($0.32\,\ts{Bl} \cdot \ts{s}^{-1}$), and the corresponding RMS value of the \mbox{lateral-position} error is $3.94\,\ts{mm}$; this \mbox{RMS-error} value represents a decrease of $80\,\%$ with respect to that obtained in open loop.

Fig.\,\ref{FIG07}(c) summarizes the experimental data of a VLEIBot\textsuperscript{+} prototype during the execution of a \mbox{$90$-degree} \mbox{left-turn} maneuver. Fig.\,\ref{FIG07}(c.i) shows a photographic composite of frames taken at intervals of $6\,\ts{s}$, of the swimming test. And, \mbox{Figs.\,\ref{FIG07}(c.ii)~and~(c.iii)} show the robot's position in the $2$D space and \mbox{lateral-position} error during the test. The measured maximum average speed and turning rate are $16.1\,\ts{mm} \cdot \ts{s}^{-1}$ ($0.35\,\ts{Bl} \cdot \ts{s}^{-1}$) and $0.25\,\ts{rad} \cdot \ts{s}^{-1}$ respectively, and the corresponding RMS value of the \mbox{lateral-position} error is $20.8\,\ts{mm}$. Fig.\,\ref{FIG07}(d) summarizes the experimental data of a VLEIBot\textsuperscript{+} prototype during the execution of a \mbox{$90$-degree} \mbox{right-turn} maneuver. Fig.\,\ref{FIG07}(d.i) shows a photographic composite of frames taken at intervals of $6\,\ts{s}$, of the swimming test. And, \mbox{Figs.\,\ref{FIG07}(d.ii)~and~(d.iii)} show the robot's position in the 2D space and \mbox{lateral-position} error during the test. The measured maximum average speed and turning rate are $10.0\,\ts{mm} \cdot \ts{s}^{-1}$ ($0.22\,\ts{Bl} \cdot \ts{s}^{-1}$) and $0.28\,\ts{rad} \cdot \ts{s}^{-1}$, respectively, and the corresponding RMS value of the \mbox{lateral-position} error is $5.45\,\ts{mm}$.

\section{Conclusions}
\label{SECTION05}
We presented the design and development of two new microrobotic swimmers: the \mbox{$45$-mg/$23$-$\ts{mm}^3$} VLEIBot and its \mbox{high-performance} evolution, the \mbox{$90$-mg/$47$-$\ts{mm}^3$} VLEIBot\textsuperscript{+}. We designed the propulsor for the VLEIBot drawing inspiration from the anguilliform swimming mode exhibited by some aquatic animals such as the European eel, \mbox{yellow-bellied} sea snake, and young ACF tadpole. The final \mbox{heuristically-optimized} propulsor design, loosely inspired by the tail of the young ACF tadpole, has a parabolic shape, a length of $26\,\ts{mm}$, and an aspect ratio of $0.41$, and produces the best results in terms of speed at an actuation frequency of $5\,\ts{Hz}$. We arrived to these parameters after performing hundreds of \mbox{tail-characterization} experiments. With the optimized propulsor, the VLEIBot can achieve forward speeds as high as \mbox{$15.1\,\ts{mm} \cdot \ts{s}^{-1}$ ($0.33\,\ts{Bl} \cdot \ts{s}^{-1}$)} at a frequency of $5\,\ts{Hz}$; however, its maneuverability is extremely limited. With optimized propulsors, the VLEIBot\textsuperscript{+} can achieve forward speeds as high as \mbox{$16.1\,\ts{mm} \cdot \ts{s}^{-1}$ ($0.35\,\ts{Bl} \cdot \ts{s}^{-1}$)} at a frequency of $5\,\ts{Hz}$; furthermore, because of its high maneuverability in the $2$D space, it can perform maneuvers at angular speeds as high as $0.28\,\ts{rad} \cdot \ts{s}^{-1}$. To our best knowledge, the VLEIBot and VLEIBot\textsuperscript{+} are the first \mbox{self-propelled} \mbox{anguilliform-inspired} microswimmers. Furthermore, these robots are two of the very few that fully exploit FSI phenomena to generate the waves that propagate along their tails to produce locomotion.

\newpage
\bibliographystyle{IEEEtran}
\bibliography{references}

\begin{thebibliography}{10}
\providecommand{\url}[1]{#1}
\csname url@samestyle\endcsname
\providecommand{\newblock}{\relax}
\providecommand{\bibinfo}[2]{#2}
\providecommand{\BIBentrySTDinterwordspacing}{\spaceskip=0pt\relax}
\providecommand{\BIBentryALTinterwordstretchfactor}{4}
\providecommand{\BIBentryALTinterwordspacing}{\spaceskip=\fontdimen2\font plus
\BIBentryALTinterwordstretchfactor\fontdimen3\font minus \fontdimen4\font\relax}
\providecommand{\BIBforeignlanguage}[2]{{%
\expandafter\ifx\csname l@#1\endcsname\relax
\typeout{** WARNING: IEEEtran.bst: No hyphenation pattern has been}%
\typeout{** loaded for the language `#1'. Using the pattern for}%
\typeout{** the default language instead.}%
\else
\language=\csname l@#1\endcsname
\fi
#2}}
\providecommand{\BIBdecl}{\relax}
\BIBdecl

\bibitem{SMALLBug_2020}
{X.-T. Nguyen}, {A. A. Calder\'on}, {A. Rigo}, {J. Z. Ge}, and {N. O. \mbox{P\'erez-Arancibia}}, ``{SMALLBug: A 30-mg Crawling Robot Driven by a High-Frequency Flexible SMA Microactuator},'' \emph{IEEE Robot. Automat. Lett.}, vol.~5, no.~4, pp. 6796--6803, Oct. 2020.

\bibitem{SMARTI_2021}
{R. M. Bena}, {X.-T. Nguyen}, {A. A. Calder\'on}, {A. Rigo}, and {N. O. \mbox{P\'erez-Arancibia}}, ``{SMARTI: A 60-mg Steerable Robot Driven by High-Frequency Shape-Memory Alloy Actuation},'' \emph{IEEE Robot. Automat. Lett.}, vol.~6, no.~4, pp. 8173--8180, Oct. 2021.

\bibitem{WaterStrider_2023}
{C. K. Trygstad}, {X.-T. Nguyen}, and {N. O. \mbox{P\'erez-Arancibia}}, ``{A New \mbox{$1$-mg} Fast Unimorph SMA-Based Actuator for Microrobotics},'' in \emph{Proc. IEEE/RSJ Int. Conf. Intell. Robots Syst. (IROS)}, Detroit, MI, USA, Oct. 2023, pp. 2693--2700.

\bibitem{PerezJBB01}
{N. O. P\'erez-Arancibia}, {K. Y. Ma}, {K. C. Galloway}, {J. D. Greenberg}, and {R. J. Wood}, ``{First controlled vertical flight of a biologically inspired microrobot},'' \emph{Bioinspir. Biomim.}, vol.~6, no.~3, Sep. 2011, {Art. no. 036009}.

\bibitem{Nestor_JINT_2015}
{N. O. P\'erez-Arancibia}, {P.-E. J. Duhamel}, {K. Y. Ma}, and {R. J. Wood}, ``{Model-Free Control of a Hovering Flapping-Wing Microrobot},'' \emph{J. Intell. Robot. Syst.}, vol.~77, no.~1, pp. 95--111, Jan. 2015.

\bibitem{BeePlus_2019}
{X. Yang}, {Y. Chen}, {L. Chang}, {A. A. Calder\'on}, and {\mbox{N. O.} \mbox{P\'erez-Arancibia}}, ``{Bee\textsuperscript{+}: A 95-mg Four-Winged Insect-Scale Flying Robot Driven by Twinned Unimorph Actuators},'' \emph{IEEE Robot. Automat. Lett.}, vol.~4, no.~4, pp. 4270--4277, Oct. 2019.

\bibitem{JINT_2022}
{R. M. Bena}, {X.-T. Nguyen}, {X. Yang}, {A. A. Calder\'on}, {Y. Chen}, and {\mbox{N. O.} \mbox{P\'erez-Arancibia}}, ``{A Multiplatform Position Control Scheme for Flying Robotic Insects},'' \emph{J. Intell. Robot. Syst.}, vol. 105, no.~1, May 2022, {Art. no. 19}.

\bibitem{BeePlusPlus_2023}
{R. M. Bena}, {X. Yang}, {A. A. Calder\'on}, and {N. O. \mbox{P\'erez-Arancibia}}, ``{High-Performance Six-DOF Flight Control of the Bee\textsuperscript{++}: An Inclined-Stroke-Plane Approach},'' \emph{IEEE Trans. Robot.}, vol.~39, no.~2, pp. 1668--1684, Apr. 2023.

\bibitem{RoBeetle_2020}
{X. Yang}, {L. Chang}, and {N. O. \mbox{P\'erez-Arancibia}}, ``{An 88-milligram insect-scale autonomous crawling robot driven by a catalytic artificial muscle},'' \emph{Sci. Robot.}, vol.~5, no.~45, Aug. 2020, {Art. no. eaba0015}.

\bibitem{Fuller2017Dampers}
{S. B. Fuller}, {Z. E. Teoh}, {P. Chirattananon}, {N. O. P\'erez-Arancibia}, {J. Greenberg}, and {R. J. Wood}, ``{Stabilizing air dampers for hovering aerial robotics: design, insect-scale flight tests, and scaling},'' \emph{Auton. Robot.}, vol.~41, no.~8, pp. 1555--1573, Dec. 2017.

\bibitem{Perez2015ModelFree}
{N. O. P\'erez-Arancibia}, {P.-E. J. Duhamel}, {K. Y. Ma}, and {R. J. Wood}, ``{Model-Free Control of a Hovering Flapping-Wing Microrobot},'' \emph{J. Intell. Robot. Syst.}, vol.~77, no.~1, pp. 95--111, Jan. 2015.

\bibitem{Perez2013ICAR}
------, ``{Model-Free Control of a Flapping-Wing Flying Microrobot},'' in \emph{Proc. 16th Int. Conf. Adv. Robot. (ICAR)}, Montevideo, Uruguay, Nov. 2013, pp. 1--8.

\bibitem{Duhamel2013OpticalFlow}
{P.-E. J. Duhamel}, {N. O. P\'erez-Arancibia}, {G. L. Barrows}, and {R. J. Wood}, ``{Biologically Inspired Optical-Flow Sensing for Altitude Control of Flapping-Wing Microrobots},'' \emph{IEEE/ASME Trans. Mechatron.}, vol.~18, no.~2, pp. 556--568, Apr. 2013.

\bibitem{Teoh2012IROS}
{Z. E. Teoh}, {S. B. Fuller}, {P. Chirarttananon}, {N. O. P\'erez-Arancibia}, {\mbox{J. D.} Greenberg}, and {R. J. Wood}, ``{A Hovering Flapping-Wing Microrobot with Altitude Control and Passive Upright Stability},'' in \emph{Proc. IEEE/RSJ Int. Conf. Intell. Robots Syst. (IROS)}, Vilamoura, Algarve, Portugal, Oct. 2012, pp. 3209--3216.

\bibitem{Duhamel2012ICRA}
{P.-E. J. Duhamel}, {N. O. P\'erez-Arancibia}, {G. L. Barrows}, and {R. J. Wood}, ``{Altitude Feedback Control of a Flapping-Wing Microrobot Using an On-Board Biologically Inspired Optical Flow Sensor},'' in \emph{Proc. IEEE Int. Conf. Robot. Autom. (ICRA)}, Saint Paul, Minnesota, USA, May 2012, pp. 4228--4235.

\bibitem{Perez2011ACC}
{N. O. P\'erez-Arancibia}, {J. P. Whitney}, and {R. J. Wood}, ``{Lift Force Control of a Flapping-Wing Microrobot},'' in \emph{Proc. Amer. Control Conf. (ACC)}, San Francisco, CA, USA, Jun. 2011, pp. 4761--4768.

\bibitem{Perez2011ROBIO}
{N. O. P\'erez-Arancibia}, {P. Chirarattananon}, {B. M. Finio}, and {R. J. Wood}, ``{Pitch-Angle Feedback Control of a Biologically Inspired Flapping-Wing Microrobot},'' in \emph{Proc. Int. Conf. Robot. Biomim. (ROBIO)}, Karon Beach, Thailand, Dec. 2011, pp. 1495--1502.

\bibitem{Perez2013Transactions}
{N. O. P\'erez-Arancibia}, {J. P. Whitney}, and {R. J. Wood}, ``{Lift Force Control of Flapping-Wing Microrobots Using Adaptive Feedforward Schemes},'' \emph{IEEE/ASME Trans. Mechatron.}, vol.~18, no.~1, pp. 155--168, Feb. 2013.

\bibitem{Palstra_EuropeanEel_2010}
{A. P. Palstra} and {G. E. E. J. M. van den Thillart}, ``{Swimming physiology of European silver eels (\textit{Anguilla anguilla} L.): energetic costs and effects on sexual maturation and reproduction},'' \emph{Fish Physiol. Biochem.}, vol.~36, no.~3, pp. 297--322, Apr. 2010.

\bibitem{Graham_SeaSnake_1986}
{J. B. Graham}, {W. R. Lowell}, {I. Rubinoff}, and {J. Motta}, ``{Surface and Subsurface Swimming of the Sea Snake \textit{Pelamis Platurus}},'' \emph{J. Exp. Biol.}, vol. 127, no.~1, pp. 27--44, Jan. 1986.

\bibitem{Roberts_XenopusLaevis_2000}
{A. Roberts}, {N. A. Hill}, and {R. Hicks}, ``{Simple Mechanisms Organise Orientation of Escape Swimming in Embryos and Hatchling Tadpoles of \textit{Xenopus Laevis}},'' \emph{J. Exp. Biol.}, vol. 203, no.~12, pp. 1869--1885, Jun. 2000.

\bibitem{Tack_Anguilliform_2021}
{N. B. Tack}, {K. T. Du Clos}, and {B. J. Gemmell}, ``{Anguilliform Locomotion across a Natural Range of Swimming Speeds},'' \emph{Fluids}, vol.~6, no.~3, Mar. 2021, {Art. no. 127}.

\bibitem{Liu_TadpoleCFD_1997}
{H. Liu}, {R. Wassersug}, and {K. Kawachi}, ``{The Three-Dimensional Hydrodynamics of Tadpole Locomotion},'' \emph{J. Exp. Biol.}, vol. 200, no.~22, pp. 2807--2819, Nov. 1997.

\bibitem{Fukuda_DistributedSMA_1990}
{T. Fukuda}, {H. Hosokai}, and {I. Kikuchi}, ``{Distributed Type of Actuators by Shape Memory Alloy and its Application to Underwater Mobile Robotic Mechanism},'' in \emph{Proc. IEEE Int. Conf. Robot. Autom. (ICRA)}, Cincinnati, OH, USA, May 1990, pp. 1316--1321.

\bibitem{Garner_SMABiomimetic_2000}
{L. J. Garner}, {L. N. Wilson}, {D. C. Lagoudas}, and {O. K. Rediniotis}, ``{Development of a shape memory alloy actuated biomimetic vehicle},'' \emph{Smart Mater. Struct.}, vol.~9, no.~5, pp. 673--683, Oct. 2000.

\bibitem{Rossi_BendingSMA_2011}
{C. Rossi}, {J. Colorado}, {W. Coral}, and {A. Barrientos}, ``{Bending continuous structures with SMAs: a novel robotic fish design},'' \emph{Bioinsp. Biomim.}, vol.~6, no.~4, Dec. 2011, {Art. no. 045005}.

\bibitem{Cho_Body_Caudal_2008}
{K.-J. Cho}, {E. Hawkes}, {C. Quinn}, and {R. J. Wood}, ``{Design, fabrication and analysis of a body-caudal fin propulsion system for a microrobotic fish},'' in \emph{Proc. IEEE Int. Conf. Robot. Autom. (ICRA)}, Pasadena, CA, USA, May 2008, pp. 706--711.

\bibitem{WANG_MicroFish_2008}
{Z. Wang}, {G. Hang}, {J. Li}, {Y. Wang}, and {K. Xiao}, ``{A micro-robot fish with embedded SMA wire actuated flexible biomimetic fin},'' \emph{Sens. Actuators A: Phys.}, vol. 144, no.~2, pp. 354--360, Jun. 2008.

\bibitem{Shi_Jellyfish_2010}
{L. Shi}, {S. Guo}, and {K. Asaka}, ``{A Novel Jellyfish-like Biomimetic Microrobot},'' in \emph{Proc. IEEE/ICME Int. Conf. Comp. Med. Eng. (CME)}, Gold Coast, QLD, Australia, Jul. 2010, pp. 277--281.

\bibitem{Byun_Helmholtz_2012}
{D. Byun}, {J. Choi}, {K. Cha}, {J.-O. Park}, and {S. Park}, ``{Swimming microrobot actuated by two pairs of Helmholtz coils system},'' \emph{Mechatronics}, vol.~21, no.~1, pp. 357--364, Feb. 2011.

\bibitem{Zhao_OrigamiSwimmer_2022}
{Q. Ze}, {S. Wu}, {J. Dai}, {S. Leanza}, {G. Ikeda}, {P. C. Yang}, {G. Iaccarino}, and {R. R. Zhao}, ``{Spinning-enabled wireless amphibious origami millirobot},'' \emph{Nat. Commun.}, vol.~13, Jun. 2022, {Art. no. 3118}.

\bibitem{Chen_MicroSwimmers_2023}
{J. Chen}, {H. Hu}, and {Y. Wang}, ``{Magnetic-driven 3D-printed biodegradable swimming microrobots},'' \emph{Smart Mater. Struct.}, vol.~32, no.~8, Jul. 2023, {Art. no. 085014}.

\bibitem{Tan_MagneticSwimmer_2023}
{L. Tan} and {D. J. Cappelleri}, ``{Design, Fabrication, and Characterization of a Helical Adaptive Multi-Material MicroRobot (HAMMR)},'' \emph{IEEE Robot. Automat. Lett.}, vol.~8, no.~3, pp. 1723--1730, Mar. 2023.

\bibitem{Seyed_Hydrogel_2011}
{S. N. Tabatabaei}, {J. Lapointe}, and {S. Martel}, ``{Shrinkable Hydrogel-Based Magnetic Microrobots for Interventions in the Vascular Network},'' \emph{Adv. Robot.}, vol.~25, no.~8, pp. 1049--1067, Apr. 2011.

\bibitem{Fusco_IntegratedMicrorobot_2014}
{S. Fusco}, {M. S. Sakar}, {S. Kennedy}, {C. Peters}, {R. Bottani}, {F. Starsich}, {A. Mao}, {G. A. Sotiriou}, {S. Pan\'e}, {S. E. Pratsinis}, {D. Mooney}, and {B. J. Nelson}, ``An integrated microrobotic platform for on-demand, targeted therapeutic interventions,'' \emph{Adv. Mater.}, vol.~26, no.~6, pp. 952--957, Feb. 2014.

\bibitem{Li_ControlledDrug_2009}
{H. Li}, {J. Tan}, and {M. Zhang}, ``{Dynamics Modeling and Analysis of a Swimming Microrobot for Controlled Drug Delivery},'' \emph{IEEE Trans. Autom. Sci. Eng.}, vol.~6, no.~2, pp. 220--227, Apr. 2009.

\bibitem{Temel_ConfinedSwimming_2015}
{F. Z. Temel} and {S. Yesilyurt}, ``{Confined swimming of bio-inspired microrobots in rectangular channels},'' \emph{Bioinspir. Biomim.}, vol.~10, no.~1, Feb. 2015, {Art. no. 016015}.

\bibitem{Palagi_SoftMagSwimmer_2011}
{S. Palagi}, {V. Pensabene}, {L. Beccai}, {B. Mazzolai}, {A. Menciassi}, and {P. Dario}, ``{Design and development of a soft magnetically-propelled swimming microrobot},'' in \emph{Proc. IEEE Int. Conf. Robot. Autom. (ICRA)}, Shanghai, China, May 2011, pp. 5109--5114.

\bibitem{Pak_HighSpeedNanowire_2011}
{O. S. Pak}, {W. Gao}, {J. Wang}, and {E. Lauga}, ``{High-speed propulsion of flexible nanowire motors: Theory and experiments},'' \emph{Soft Matter}, vol.~7, no.~18, Sep. 2011, {Art. no. 8169}.

\bibitem{DreyfusMicroArtSwimmer_2005}
{R. Dreyfus}, {J. Baudry}, {M. L. Roper}, {M. Fermigier}, {H. A. Stone}, and {J. Bibette}, ``{Microscopic artificial swimmers},'' \emph{Nature}, vol. 437, no.~6, pp. 862--865, Oct. 2005.

\bibitem{GaoFlexibleNanowireMotor_2010}
{W. Gao}, {S. Sattayasamitsathit}, {K. M. Manesh}, {D. Weihs}, and {J. Wang}, ``{Magnetically Powered Flexible Metal Nanowire Motors},'' \emph{J. Am. Chem. Soc.}, vol. 132, no.~41, pp. 14\,403--14\,405, Sep. 2010.

\bibitem{Xu2013HelicalMicroswimmers}
{T. Xu}, {G. Hwang}, {N. Andreff}, and {S. R\'egnier}, ``{The rotational propulsion characteristics of scaled-up helical microswimmers with different heads and magnetic positioning},'' in \emph{Proc. IEEE/ASME Int. Conf. Adv. Intell. Mechatron. (AIM)}, Wollongong, NSW, Australia, Jul. 2013, pp. 1114--1120.

\bibitem{Ghosh2009ControlledPropulsion}
{A. Ghosh} and {P. Fischer}, ``{Controlled Propulsion of Artificial Magnetic Nanostructured Propellers},'' \emph{Nano Lett.}, vol.~9, no.~6, pp. 2243--2245, May 2009.

\bibitem{Sing2009CollidalWalkers}
{C. E. Sing}, {L. Schmid}, {M. F. Schneider}, and {A. Alexander-Katz}, ``{Controlled surface-induced flows from the motion of self-assembled colloidal walkers},'' \emph{Proc. Natl. Acad. Sci.}, vol. 107, no.~2, pp. 535--540, Dec. 2009.

\bibitem{Zhang2010RotatingNickleNanowires}
{L. Zhang}, {T. Petit}, {Y. Lu}, {B. E. Kratochvil}, {K. E. Peyer}, {R. Pei}, {J. Lou}, and {B. J. Nelson}, ``{Controlled Propulsion and Cargo Transport of Rotating Nickle Nanowires near a Patterned Solid Surface},'' \emph{ACS Nano}, vol.~4, no.~10, pp. 6228--6234, Sep. 2010.

\bibitem{Liu2010WirelessSwimmingMicrorobot}
{W. Liu}, {X. Jia}, {F. Wang}, and {Z. Jia}, ``{An in-pipe wireless swimming microrobot driven by giant magnetostrictive film},'' \emph{Sens. Actuators A: Phys.}, vol. 160, no.~2, pp. 101--108, May 2010.

\bibitem{Tierno2008ColloidalMicroswimmers}
{P. Tierno}, {R. Golestanian}, {I. Pagonabarraga}, and {F. Sagu\'es}, ``{Magnetically Actuated Colloidal Microswimmers},'' \emph{J. Phys. Chem. B}, vol. 112, no.~51, pp. 16\,525--16\,528, Nov. 2008.

\bibitem{Floyd2008UntetheredMagneticallyActuated}
{S. Floyd}, {C. Pawashe}, and {M. Sitti}, ``{An Untethered Magnetically Actuated Micro-Robot Capable of Motion on Arbitrary Surfaces},'' in \emph{Proc. IEEE/ASME Int. Conf. Robot. Autom. (ICRA)}, Pasadena, CA. USA, May. 2008, pp. 419--424.

\bibitem{Zhao2021PZTFrog}
{Q. Zhao}, {S. Liu}, {J. Chen}, {G. He}, {J. Di}, {L. Zhao}, {T. Su}, {M. Zhang}, and {Z. Hou}, ``{Fast-moving piezoelectric micro-robotic fish with double caudal fins},'' \emph{Robot. Auton. Syst.}, vol. 140, Jun. 2021, {Art. no 103733}.

\bibitem{Li2023PZT}
{K. Li}, {X. Zhou}, {Y. Liu}, {J. Sun}, {X. Tian}, {H. Zheng}, {. Zhang}, {J. Deng}, {J. Liu}, {W. Chen}, and {J. Zhao}, ``A 5 cm-scale piezoelectric jetting agile underwater robot,'' \emph{Adv. Intell. Syst.}, vol.~5, no.~4, Apr. 2023, {Art. no 2200262}.

\bibitem{song2007PZTWaterStrider}
{Y. S. Song} and {M. Sitti}, ``{Surface-Tension-Driven Biologically Inspired Water Strider Robots: Theory and Experiments},'' \emph{IEEE Trans. Robot.}, vol.~23, no.~3, pp. 578--589, Jun. 2007.

\bibitem{Chen2018HybridTerrestrialAquaticMicrorobot}
{Y. Chen}, {N. Doshi}, {B. Goldberg}, {H. Wang}, and {R. J. Wood}, ``{Controllable water surface to underwater transition through electrowetting in a hybrid terrestrial-aquatic microrobot},'' \emph{Nature Commu.}, vol.~9, Jun. 2018, art. no. 2495.

\bibitem{Deng2005OscillatingFinPropulsor}
{X. Deng} and {S. Avadhanula}, ``{Biomimetic Micro Underwater Vehicle with Oscillating Fin Propulsion: System Design and Force Measurement},'' in \emph{Proc. IEEE/ASME Int. Conf. Robot. Autom. (ICRA)}, Barcelona, Spain, Apr. 2005, pp. 3312--3317.

\bibitem{Ming2009PiezoelectricFiberComposite}
{A. Ming}, {S. Park}, {Y. Nagata}, and {M. Shimojo}, ``{Development of Underwater Robots using Piezoelectric Fiber Composite},'' in \emph{Proc. IEEE/ASME Int. Conf. Robot. Autom. (ICRA)}, Kobe, Japan, May 2009, pp. 3821--3826.

\bibitem{Fukuda1994MicroMobileRobots}
{T. Fukuda}, {A. Kawamoto}, {F. Arai}, and {H. Matsuura}, ``{Mechanism and Swimming Experiment of Micro Mobile Robot in Water},'' in \emph{Proc. IEEE/ASME Int. Conf. Robot. Autom. (ICRA)}, San Diego, CA, USA, May 1994, pp. 814--819.

\bibitem{Kim_2005}
{B. Kim}, {D.-H. Kim}, {J. Jung}, and {J.-O. Park}, ``{A biomimetic undulatory tadpole robot using ionic polymer–metal composite actuators},'' \emph{Smart Mater. Struct.}, vol.~14, no.~6, pp. 1579--1585, Dec. 2005.

\bibitem{Kamamichi2006SwimmingSnake}
{N. Kamamichi}, {M. Yamakita}, {K. Asaka}, and {Z.-W. Luo}, ``{A Snake-like Swimming Robot Using IPMC Actuator/Sensor},'' in \emph{Proc. IEEE/ASME Int. Conf. Robot. Autom. (ICRA)}, Orlando, Florida, May 2006, pp. 1812--1817.

\bibitem{Guo2003FishLikeMicrorobot}
{S. Guo}, {T. Fukuda}, and {K. Asaka}, ``{A New Type of Fish-Like Underwater Microrobot},'' \emph{IEEE/ASME Trans. Robot.}, vol.~8, no.~1, pp. 136--141, Mar. 2003.

\bibitem{PurcellAJP_1977}
{E. M. Purcell}, ``{Life at low Reynolds number},'' \emph{Am. J. Phys.}, vol.~45, no.~1, pp. 3--11, Jan. 1977.

\bibitem{Finio2011SystemID}
{B. M. Finio}, {N. O. P\'erez-Arancibia}, and {R. J. Wood}, ``{System identification and linear time-invariant modeling of an insect-sized flapping-wing micro air vehicle},'' in \emph{Proc. IEEE Int. Conf. Intell. Robots Syst. (IROS)}, San Francisco, CA. USA, Sep. 2011, pp. 1107--1114.

\end{thebibliography}
\balance
\end{document}